\documentclass[3p,times,procedia]{elsarticle}
\usepackage{nupha_ecrc}

\volume{00}

\firstpage{1}

\journalname{Nuclear Physics A}

\runauth{}

\jid{nupha}

\jnltitlelogo{Nuclear Physics A}

\usepackage{amssymb}

 \usepackage{lineno}

\usepackage[figuresright]{rotating}

\begin{document}

\begin{frontmatter}


\author{Jason Bryslawskyj for the PHENIX collaboration}
\address{University of California, Riverside, CA 92507}

\dochead{XXVIIth International Conference on Ultrarelativistic Nucleus-Nucleus Collisions\\ (Quark Matter 2018)}

\title{PHENIX study of the initial state with forward hadron measurements in 200 GeV p(d)+A and $^{3}$He$+$Au collisions}


\author{}

\address{}

\begin{abstract}
  Forward hadron measurements in p(d)+A provide a signal to study nuclear shadowing, initial state energy loss and/or gluon saturation effects as a function of rapidity, centrality and energy. High $p_T$ identified $\pi^{0}$ measurements are an essential first step toward measuring prompt photon production. The $\pi^{0}$ measurements are enabled by the PHENIX MPC-EX detector, a Si-W preshower detector located in front of the Muon Piston Calorimeter (MPC), expanding the neutral pion reconstruction capabilities in the rapidity range $3.1< \eta <3.8$ out to high energies, $E < 80$ GeV. Previous PHENIX measurements of punch-through charged hadrons in the muon arms in the rapidity range $1.4< \vert \eta \vert <2.2$ were significantly improved through the capability of the forward silicon vertex detector (FVTX) to determine the transverse momentum and rapidity with high precision and reject background from secondary hadrons.

PHENIX collected d+Au data with the MPC-EX in the 2016 run at 
$\sqrt{s_{NN}} =$ 200, 62, 39 and 19.6 GeV; and p+p and p+Au(Al) data with the FVTX in 2015 at 200 GeV. In this talk, we will present first results for high $p_T$ $\pi^{0}$ production from the $\sqrt{s_{NN}}$ = 200 GeV dataset, the status of the prompt photon measurement, as well as charged hadron nuclear modification factors in p+Au(Al) and $^{3}$He$+$Au.
\end{abstract}


\end{frontmatter}


\section{Introduction}
\label{}
Asymmetric collision systems provide unique insight in to the nature of strongly interacting matter. New PHENIX measurements are presented for forward and backward charged hadron suppression in 200 GeV p+Au and p+Al. Small-$x$ regions of the heavy nucleus can be probed at forward rapidity, whereas large-$x$ regions can be examined at backwards rapidity. The forward silicon vertex detector (FVTX) was leveraged to improve the reconstruction of punch-through charged hadrons in the PHENIX muon arm spectrometers. The FVTX consists of 4 layers of silicon tracking planes placed close to the collision vertex. Background particles which do not originate from the collision vertex, such as secondary hadrons are rejected by requiring hits in the FVTX. By obtaining precision position measurements near the collision vertex, the FVTX also allows measurement of particle (pseudo)rapidity to high precision. Muon contribution in the measured transverse momentum range ($p_T>1.5$) is estimated to be less than $5\%$, based on Geant 4 simulation of hadron cocktails with realistic $p_T$ and pseudorapidity ($\eta$) distributions.

\begin{figure*}[h!]
  \centering
  \includegraphics[width=1.05\textwidth]{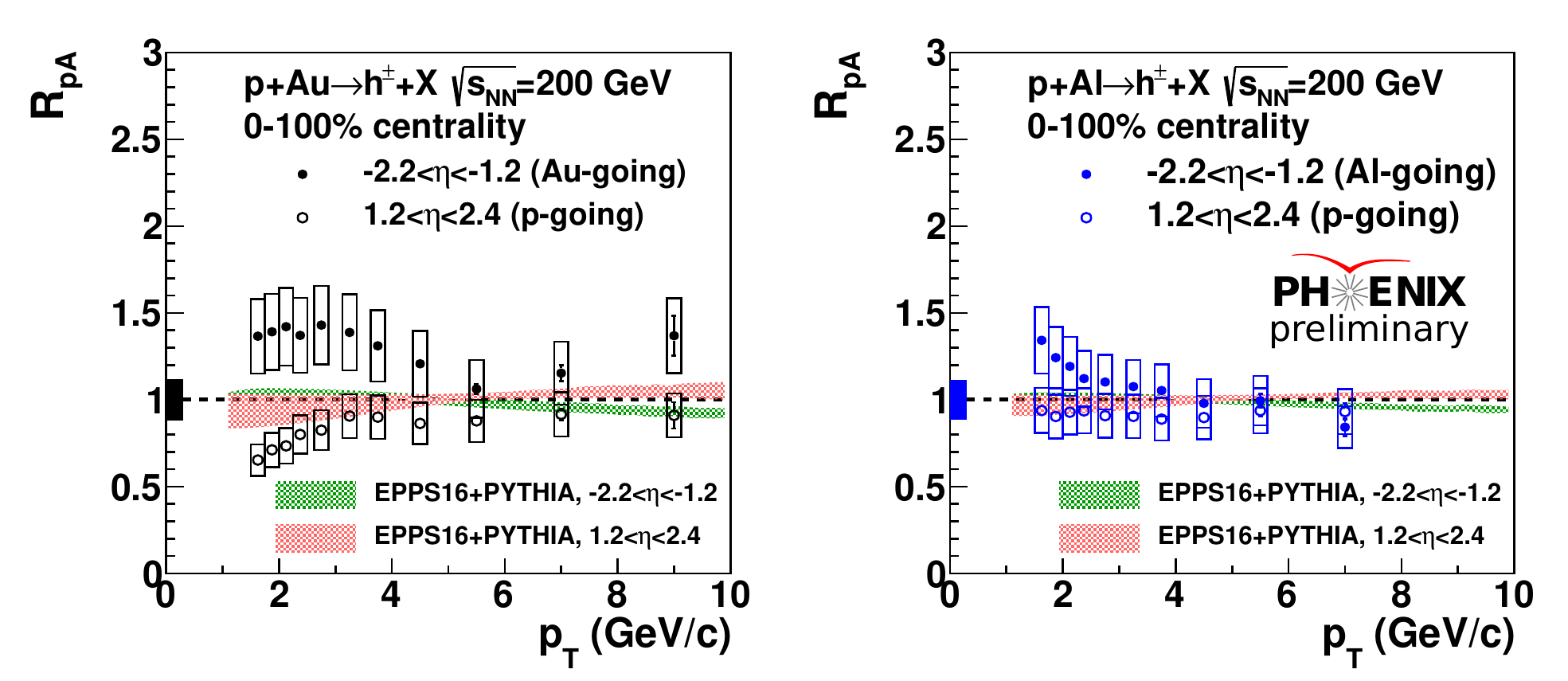}
  \caption{Nuclear modification factor ($R_{pA}$) of charged hadrons as a function of transverse momentum ($p_T$) in 200 GeV p+Au (left) and 200 GeV p+Al (right). Forward hadrons are shown as open circles and backward hadrons as closed. Data is compared to an $R_{pA}$ calculated from EPPS16\cite{Eskola:2016oht}. We define forward (pseudo)rapidity as the p-going direction and backwards in the A-going direction.  \label{fig:RpAvspT} }
\end{figure*}

\section{Results and Discussion}
Single particle modification is quantified with the nuclear modification factor $R_{pA}$:
\begin{equation}
  R_{pA} = \frac{(1/N_{pA}^{\rm evt})  d^{2}N_{pA}/dp_T dy}{\langle T_{\tiny pA} \rangle \times  d^{2}\sigma_{pp}/dp_T dy },
\end{equation}

\noindent where $\sigma_{pp}$ is the p+p cross section for the single hadron process and $\langle T_{pA}\rangle$ is the nuclear thickness function. The nuclear modification factor ($R_{pA}$) of charged hadrons as a function of transverse momentum ($p_T$) is shown on the left hand plot of Figure \ref{fig:RpAvspT} for 200 GeV p+Au compared to 200 GeV p+Al on the right hand side. Forward hadrons in the pseudorapidity range $1.2<\eta<2.4$ are plotted as filled circles and backward hadrons $(-2.2<\eta<-1.2)$ as open circles. Both are compared to nuclear modification factors calculated using the nuclear parton distribution function (nPDF) EPPS16\cite{Eskola:2016oht}. The comparison $R_{pA}$ is calculated from the EPPS16 nPDF using PYTHIA, specifically the Pythia8 MB (SoftQCD:inelastic) tune using the CT10 pp-pdf.

\begin{figure*}[h!]
  \centering
  \includegraphics[width=1.05\textwidth]{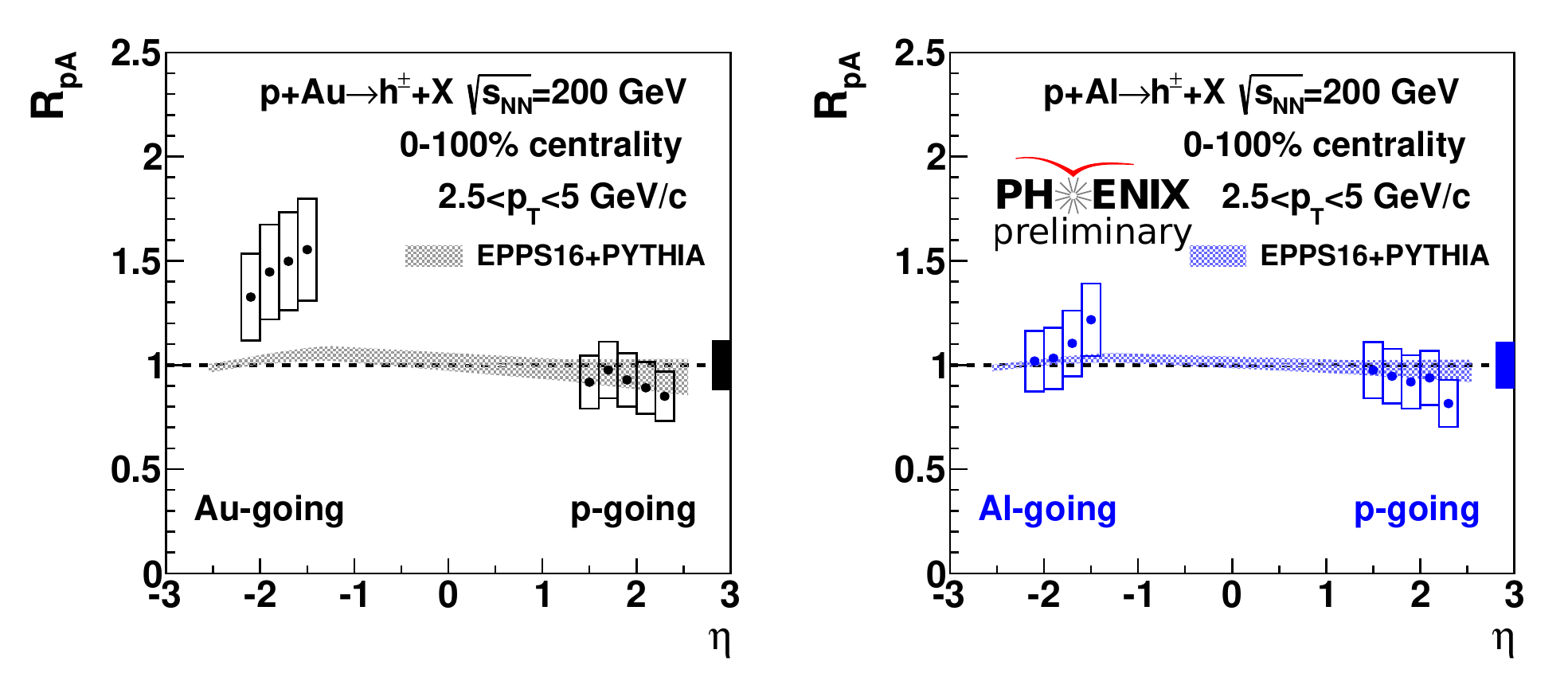}
  \caption{Nuclear modification factor ($R_{pA}$) as a function of pseudorapidity $\eta$ of charged hadrons in 200 GeV p+Au (left) and 200 GeV p+Al (right). Forward hadrons are shown as open circles and backward hadrons as closed. Data is compared to $R_{pA}$ calculated using EPPS16\cite{Eskola:2016oht}.}
  \label{fig:RpAvseta}
\end{figure*}

As can be seen in Figure \ref{fig:RpAvspT}, forward charged hadrons are suppressed, especially at low $p_T$. They are suppressed to a greater degree in the larger system p+Au than in p+Al. The suppression of forward hadrons is consistent with an $R_{pA}$ calculated using EPPS16 \cite{Eskola:2016oht}. Backward hadrons, on the other hand, are enhanced in p+Au and to a lesser extent in p+Al. These results are consistent with previous PHENIX measurements of forward and backward heavy flavor leptons as well as $\phi$ mesons in 200 GeV d+Au\cite{Adare:2017bhp}.

\begin{figure*}[h!]
  \centering
  \includegraphics[width=0.96\textwidth]{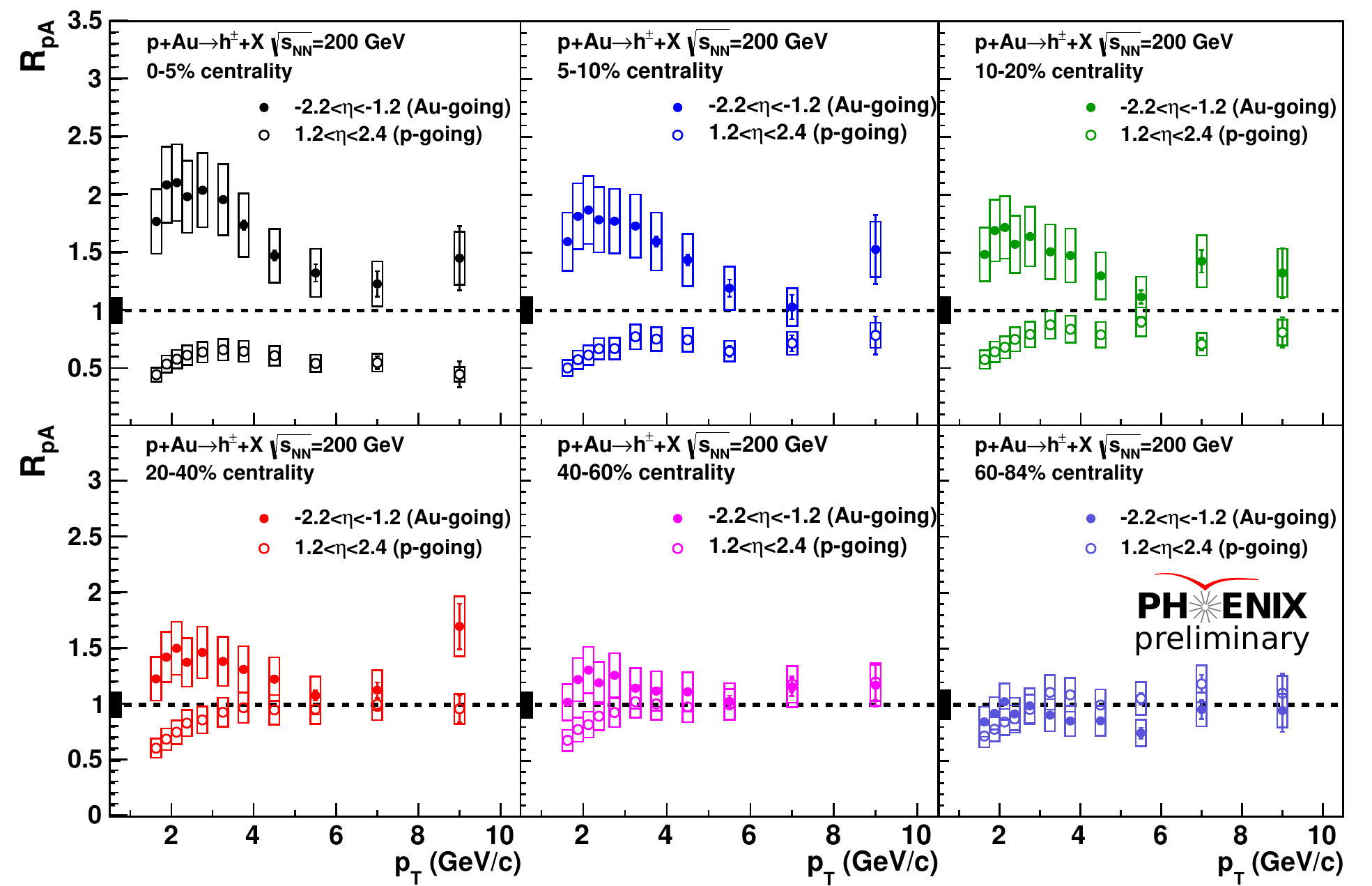}
  \caption{Charged hadron nuclear modification factor ($R_{dAu}$) in 200 GeV p+Au as a function of transverse momentum ($p_T$) plotted for various centralities. Both the backward enhancement as well as the forward suppression appear to depend strongly on centrality. }
  \label{fig:RdA_cent}
\end{figure*}

The enhancement of backward hadrons is greater in magnitude than the calculated $R_{pA}$ from EPPS16. One can see this especially well in Figure ~\ref{fig:RpAvseta}, where $R_{pA}$ is plotted as a function of pseudorapidity $\eta$. This indicates the existence of effects beyond those explained by the nuclear parton distribution function alone. This enhancement  has several possible explanations. It may be due to transverse momentum broadening from incoherent multiple scattering of partons in the larger nucleus\cite{Kang:2013ufa},\cite{Kang:2014hha}. Another possibility, is a backward enhancement caused by radial flow, as discussed for heavy-flavor quarks in \cite{Beraudo:2015wsd} and consistent with the blast-wave model fit by the STAR experiment to light-flavor spectra\cite{Abelev:2008ab}.

\begin{figure*}[h!]
  \centering
  \includegraphics[width=0.7\textwidth]{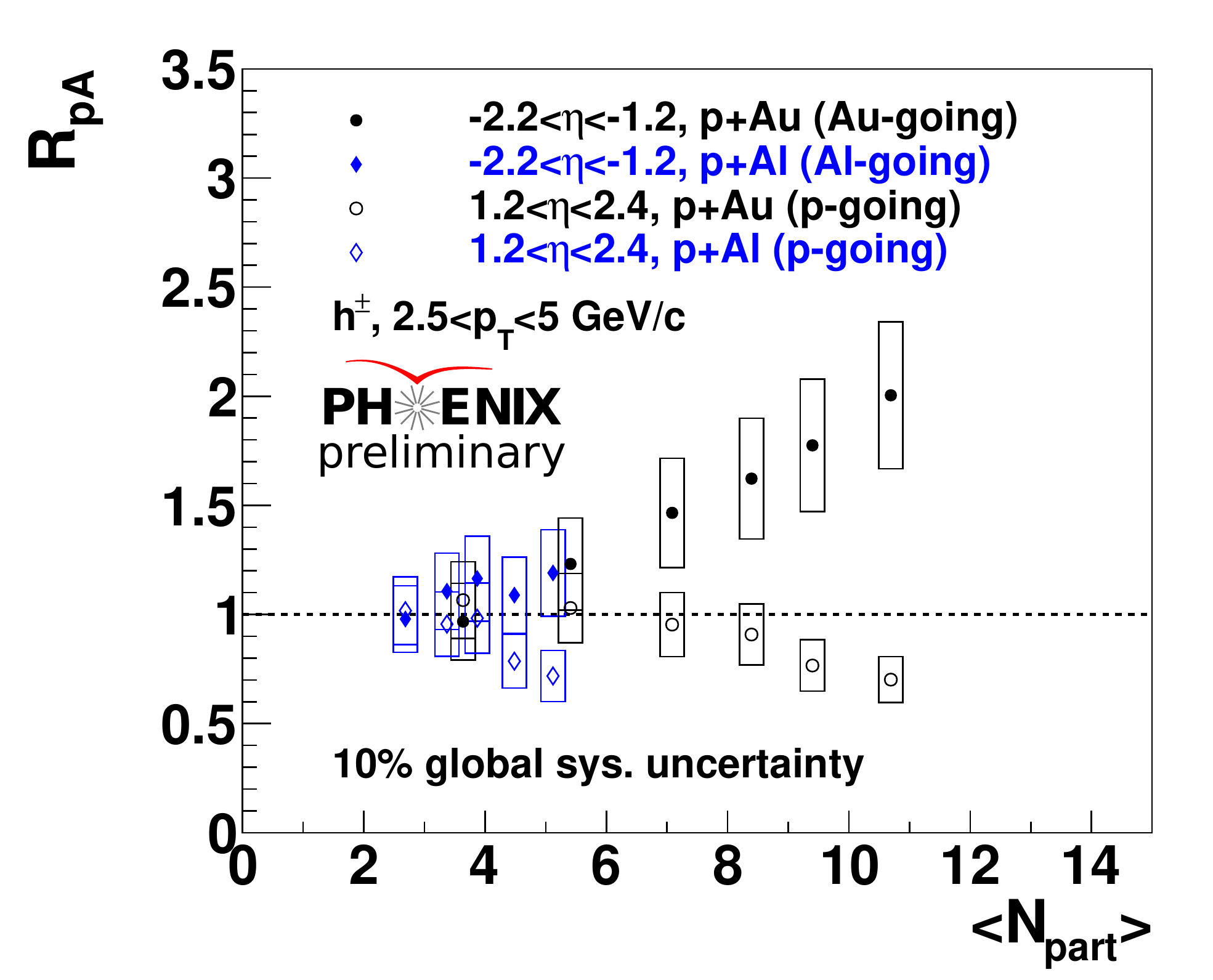}
  \caption{Charged hadron nuclear modification factor ($R_{dAu}$) as a function of the number of participants $\langle N_{\mbox{\tiny part}} \rangle $, plotted for both 200 GeV p+Au and p+Al. The backward enhancement of charged hadrons appears to scale with $\langle N_{\mbox{\tiny part}} \rangle $.}
  \label{fig:RdA_Npart}
\end{figure*}

Examining $R_{pA}$ as a function of centrality, shows a clear centrality dependence of both the backward enhancement as well as the forward suppression of charged hadrons. Centrality is determined based on the charge distribution in beam-beam counters located within a pseudorapidity range of $-3.9<\eta<-3.1$. $R_{pA}$ vs. $p_T$ is shown for various centrality bins in Figure \ref{fig:RdA_cent} for 200 GeV p+Au. Both effects are most apparent in the most central bin, 0-5$\%$. Similarly, $R_{pA}$ is plotted as a function of the number of participants ($\langle N_{\mbox{\tiny part}} \rangle $) (calculated from Glauber Monte Carlo) in Figure \ref{fig:RdA_Npart} for both p+Au and p+Al. The enhancement of backward hadrons scales with $\langle N_{\mbox{\tiny part}} \rangle $ across the two different collision systems.

\section{Conclusion and Future Measurements}
New measurements of forward and backward charged hadrons from PHENIX in 200 GeV p+Au and p+Al are presented. Forward p-going hadrons are suppressed as expected from a variety of different effects. Backward Au-going hadrons are enhanced, perhaps due to incoherent parton multiple scattering \cite{Kang:2013ufa},\cite{Kang:2014hha} or due to radial flow\cite{Beraudo:2015wsd},\cite{Abelev:2008ab}. Both effects appear larger in p+Au and in more central collisions. The backward enhancement, in fact, scales with the number of participants $\langle N_{\mbox{\tiny part}} \rangle $.

In addition to the presented measurements at high rapidity ($1.2 < |\eta| \lesssim 2.4$), PHENIX looks forward to new measurements of hadrons at even higher rapidity. Using the MPC-EX silicon preshower upgrade to the Muon Piston Calorimeter, we will be able to measure high-$p_T$ $\pi^0$s and direct photons at very high rapidity ($3.1<|\eta| < 3.8$) and at low-$x$. Measuring low-$x$ direct photons will allow direct access to the underlying gluon distribution and provide insight into gluon suppression at low-$x$ and high centrality.





\bibliographystyle{elsarticle-num}
\bibliography{QM_2018_Bryslawskyj.bib}







\end{document}